\documentclass[fleqn,usenatbib]{mnras}

\usepackage{newtxtext,newtxmath}

\usepackage[T1]{fontenc}

\usepackage{graphicx}	
\usepackage{amsmath}	

\usepackage{amssymb}	

\usepackage{xcolor}
\definecolor{red}{RGB}{227,37,107}
\definecolor{peri}{RGB}{10,11,216}
\definecolor{smar}{RGB}{255, 117, 24}

\title[Occultation in 21\;cm Cosmology]{Quantifying the Impact of Lunar and Planetary Occultation on Experimental Global 21 cm Cosmology}

\author[J. H. N. Pattison et al.]{
Joe H. N. Pattison,$^{1,2}$\thanks{E-mail: jhnp2@cam.ac.uk}
Dominic J. Anstey,$^{1,2}$\thanks{E-mail: da401@cam.ac.uk}
Eloy de Lera Acedo$^{1,2}$\thanks{E-mail: ed330@cam.ac.uk}
\\
$^{1}$Astrophysics Group, Cavendish Laboratory, J.J. Thomson Avenue, Cambridge, CB3 0HE, UK\\
$^{2}$Kavli Institue for Cosmology, Madingley Road, Cambridge, CB3 0HA, UK\\
}

\date{Accepted XXX. Received YYY; in original form ZZZ}

\pubyear{\the\year{}}

\begin{document}
\label{firstpage}
\pagerange{\pageref{firstpage}--\pageref{lastpage}}
\maketitle
\defcitealias{ptolemy_almagest}{Ptolemy, c.~150}
\begin{abstract}

The global 21\,cm signal from the hyperfine transition of cosmic atomic hydrogen is theorised to track the state of the early Universe via the analysis of its absorption and emission with respect to the radio background.
Detecting this signal has been a challenge for astronomers since it was first postulated due to the presence of strong galactic foregrounds obfuscating the transition.
Forward modelling techniques that aim to simulate and then remove these foregrounds have been suggested as a workaround to this problem.
This technique, however, requires a precise and accurate understanding of the foregrounds in question.
As we move into the next major lunar standstill, the moon will be able to occult high power areas of the sky in ways that are unaccounted for by maps used to simulate these foregrounds, and thereby disrupt signal recovery.
We show that in toy cases an occultation from the moon, or other proximate object, leading to a mismatch in our expected and actual foregrounds of 15 parts per million increases the error in signal recovery of up to 20\%.
We show that this level of disruption can happen when the moon is found towards the centre of the galaxy and is amplified when this alignment happens at high directivity regions of the antenna beam, causing a disruption of up to 180 parts per million, leading to a signal recovery error of 115\%.
This allows us to identify lunar alignment scenarios that should be actively avoided to preserve signal fidelity.
We also demonstrate that a body with a smaller apparent size than the moon, such as Venus, is unlikely to cause any signal disruption due to occultation, giving a base map error of <2 parts per million.

\end{abstract}




\section{Introduction}
21\;cm cosmology aims to describe the early Universe via the observation of the redshifted hyperfine spin transition of cosmic hydrogen.
The spin temperature, \(T_s\), of this transition, that is, the ratio of the population of hydrogen atoms in the 1S singlet and triplet levels, is impacted by: emission and absorption of photons to and from the radio background, collisions with other atoms, ions, or leptons, and Ly\(\alpha\) scattering \citep{Pritchard201221cmCentury}.
This \(T_s\) is determined from measurements of the brightness temperature of the 21\,cm signal (assuming accurate description of the radio background).
Global 21\,cm cosmology is a field that aims to find the sky averaged brightness temperature of the 21\,cm signal from cosmic hydrogen across a range of frequencies using a single antenna.
This will then determine \(T_s\) across time as the signal redshifts, allowing cosmologists to track the state of the early Universe, giving one of the more salient probes of cosmic dawn and the epoch of reionisation.

Discerning this brightness temperature from all other sources, however, is a challenging task.
With an intensity 5 orders of magnitude lower than the foregrounds, the 21\;cm signal is easily drowned out by non-cosmological sources, such as galactic synchrotron radiation, ionospheric activity \citep{Shen2021QuantifyingObservations}, and the horizon around the radiometer \citep{Bassett2021LostAnalysis, Pattison2024ModellingForegrounds}.
Thus the ability of a physically motivated foreground method to recover the 21\;cm line is highly sensitive to the accuracy of the foregrounds used in the model.
The sensitivity of signal recovery with respect to the foreground model is further complicated by the effects of antenna chromaticity \citep{Anstey2020AExperiments}, where, with changing frequencies, different areas of the sky will inject more power into the antenna than others.

There are many bodies in the celestial sphere that sit in front of the foreground maps used for 21\,cm cosmology that have non-sidereal motion.
These bodies occulting areas of the sky will lead to errors in the foreground maps used for signal recovery.
The largest of these bodies are the sun and moon, both of which have an apparent angular size of \(\sim\)0.5 degrees across the sky.
Observation windows with the sun overhead are generally avoided in 21\,cm cosmology as the solar impact on the ionosphere \citep{Hayes2021SolarIonosphere} exacerbates problems with modelling that region of the atmosphere beyond reasonable levels.
While we are able to choose to not observe while the sun is up, avoiding both the sun and moon - which are equal in apparent angular size - would dramatically reduce our possible observing windows.
Instead, it is important to more carefully quantify the level to which occultation may be a problem, and avoid observing windows where it becomes one.

\begin{figure*}
    \centering
    \includegraphics[width = \textwidth]{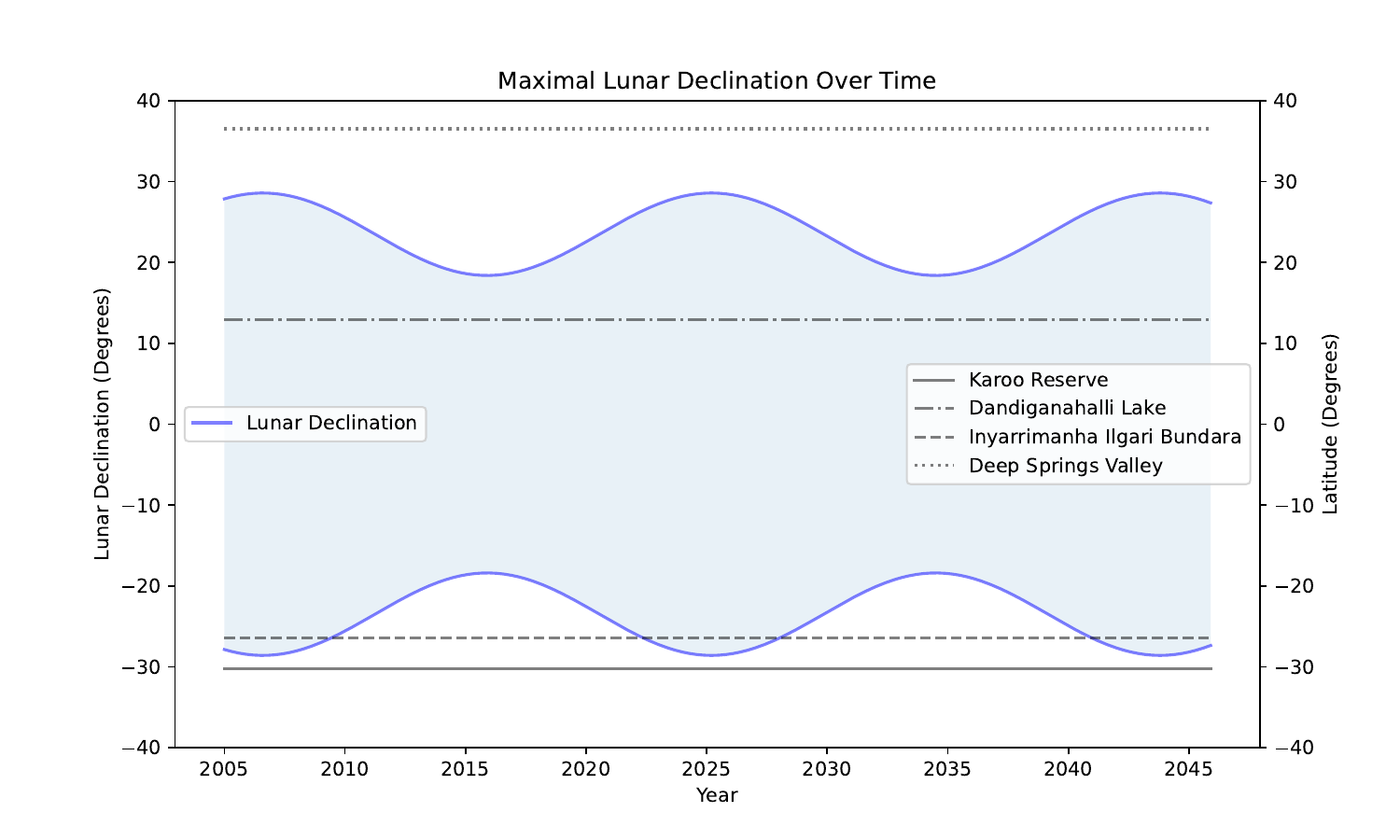}
    \caption{A comparison of the locations of four global 21\;cm experiments with the maximal declination of the moon over a period of approximately 40 years. Maximal lunar declination is shown as a continuous blue line, and the shaded blue area describes all possible lunar declinations over the course of a tropical month. Site latitudes are indicated by a grey line with varying styles. For this work we chose to look at the MIST Deep Springs Valley site as it the closest of the MIST sites to the equator, though it should be noted that MIST is an experiment designed to be mobile and is not fixed in one locale.}
    \label{fig:Declination}
\end{figure*}

The lunar orbital plane is offset \(5.1^\circ\) from the ecliptic, which is misaligned from the equatorial plane by \(23.5^\circ\).
The 18.6 year cycle of the precession of the lunar nodes around the ecliptic \citep{Folkner2014TheDE431b} means that the maximum and minimum declination of the moon will vary over time.
The least of this variation, or a minor lunar standstill, means a lunar declination of \(\pm18.1^\circ\), and the  greatest variation here will mean the moon is able to reach declinations of \(\pm28.7^\circ\) in what is known as a major lunar standstill - the last of which was in 2006, with another in 2025.
As we move into the major lunar standstill, the moon at a declination of \(\pm28.7^\circ\) is able to reach a similar declination to the galactic bulge (\(29^\circ\)), a major source of power in the sky, the occultation of which would cause great disruption to the foreground power output.
This is especially problematic for experiments like REACH \citep{deLeraAcedo2022The7.528} and EDGES \citep{Bowman2018AnSpectrum}, which sit at a latitude where the galaxy will also reach zenith, thereby compounding the impact of a misunderstood foreground region in high directivity areas of the antenna beam. 
This is shown in Figure \ref{fig:Declination}, where we compare the latitude of four global 21\;cm experiments across the globe to the maximal declination of the moon over a period of 40 years.
For experiments near the equator or in the northern hemisphere, like SARAS \citep{Singh2017SARASSignal} or MIST \citep{Monsalve2024MapperOverview} the moon will still be able to find its way into the galactic centre, or near the beam zenith, but will not do so at the same time due to the locality of the observing sites.

The impact of occultation by the moon, or other bodies, on global 21\,cm cosmology is poorly understood.
Some experiments have investigated this impact, with EDGES noting a 0.05\,K increase when the moon is more than 45 degrees over the horizon, though there is no work done here to quantify the impact of the moon being over the horizon in different regions of the sky.
Other experiments, like SARAS-3, have not aimed to quantify this disturbance, or avoid it when observing.

In this paper we aim to quantify the impact of lunar occultation on a range of global 21\;cm simulations at different latitudes and times across the world using the data analysis pipeline designed for the REACH \citep{Anstey2020AExperiments}.
We aim to disseminate the extent to which the moon and occulting bodies with a smaller apparent angular size are able to impact the ability of a forward modelled pipeline to recover the 21\,cm signal, and the observing windows for which one should be especially cautious.

Section \ref{sec:methods} details how we describe the antenna beams, foreground maps, occultation due to the moon, and the data analysis tools we use; Section \ref{sec:Results} discusses the impact of occultation in static and realistic simulations, as well as investigating the impact of bodies with an apparent angular radius smaller than that of the moon; Section \ref{sec:conc} concludes and suggests steps to mitigate impact from occultation.

\section{Methods}
\label{sec:methods}

Here we give an overview of the methods used to model and analyse occultation.
In Section \ref{sec:beams} we briefly detail the antenna beams used in these simulations.
Sections and \ref{sec:modelling} and \ref{sec:moonmodel} explain how we go about modelling the galactic foregrounds and then the occulting bodies used this this work.
Section \ref{sec:dataanalysis} discusses the data analysis tools used to recovery the redshifted 21\,cm signal in our simulations.

\subsection{Antenna Beams}
\label{sec:beams}

\begin{figure}
    \centering
    \includegraphics[width=\linewidth]{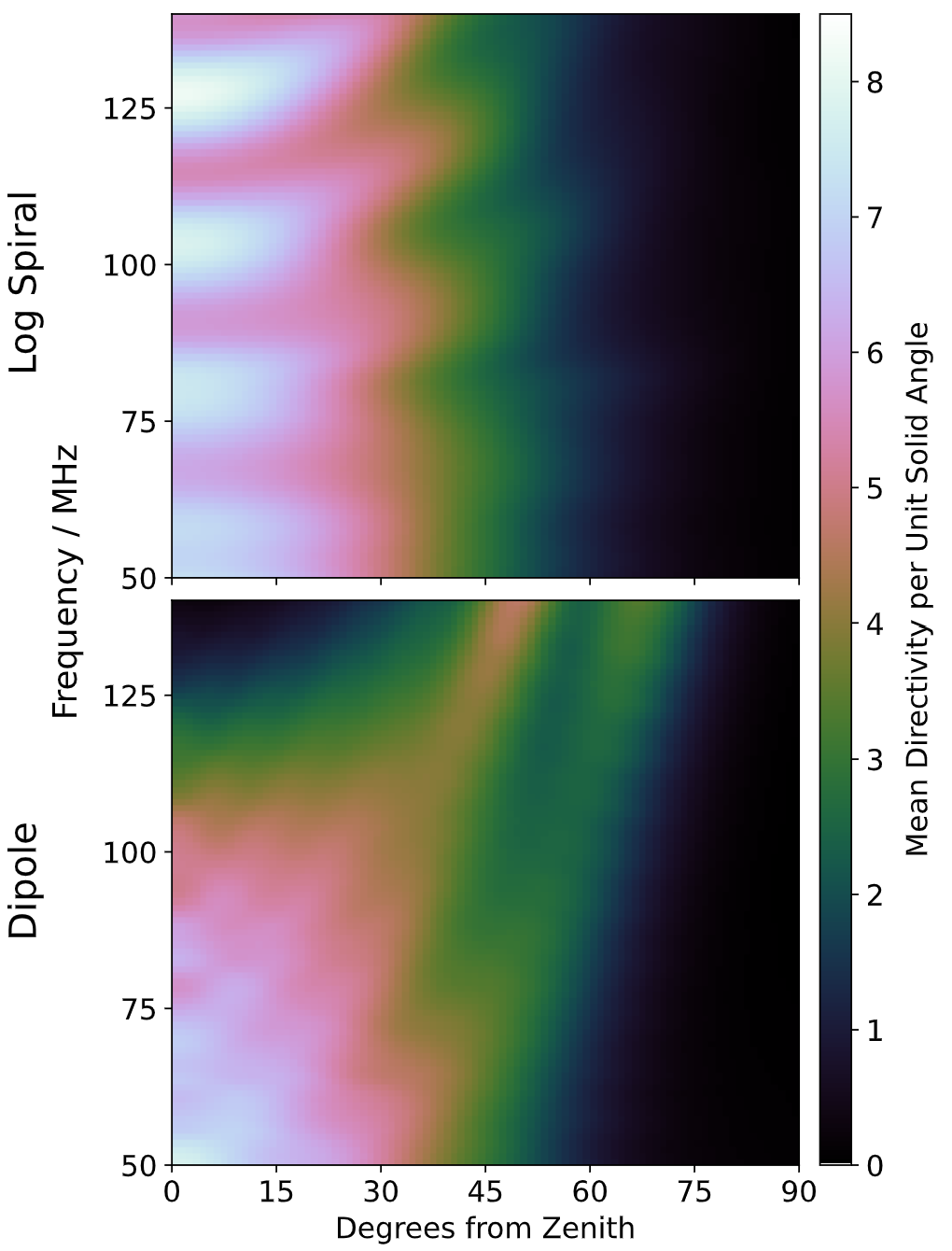}
    \caption{Waterfall plot showing the chromaticity of a vacuum backed 6\,m conical log spiral antenna (top) and vacuum backed REACH-like dipole antenna (bottom). The `cubehelix' \citep{Green2011AImages} colourmap shows the mean directivity per unit solid angle at given angle from zenith for all values of \(\phi\).}
    \label{fig:directive}
\end{figure}

In this work when investigating lunar occultation and its impact on signal recovery we use two beams designed for the REACH experiment - a vacuum backed REACH-like dipole \citep{Cumner2022RadioCase} and 6\,m conical log spiral antenna, the directivities of which are shown in Figure \ref{fig:directive}.
These beams were created using CST Studio Suite\textregistered.

Precise and accurate modelling of the beam is necessary for any physically motivated recovery of the 21\,cm signal; an unmodelled dent in a ground plane, or variation in the soil under the antenna may mean that accurate signal recovery becomes impossible \citep{Cumner2024TheExperiment, Pattison2025GlobalConditions}.
In this work we assume a perfect modelling of the antenna between our data simulations and foreground models.
This assumption will allow us to investigate the interplay between the error in the sky map caused by the moon, and the subsequent beam-weighted error caused by the combination of the beam and the occulted sky.
We therefore aim to look into the extent to which position of the occultation with respect to the beam versus simply the effect that the occultation has on the sky itself.

\subsection{Sky Modelling}
\label{sec:modelling}
To create a model of the sky we follow \citet{Pattison2024ModellingForegrounds} -- a more thorough description of this process can be found there.

We generate a map of the varying spectral indices across the sky by generating an instance of the 2008 Global Sky Model (GSM)\citep{deOliveira-Costa2008AGHz} at 408\,MHz and transforming it onto a corresponding map at 230\,MHz in an attempt to avoid the redshifted cosmological 21\,cm signal.
This map is then rotated according to the location and time of observation, giving a representation of what an antenna would see.

We then generate the horizon surrounding the antennae at each location using the \textsc{shapes} algorithm \citep{Bassett2021LostAnalysis}.
This is used to mask the sky models previously made, and the horizon is treated as outputting a blackbody  power of 300\,K, and reflecting the sky, and its own power with a reflection coefficient of 0.6.
Next we inject a theoretical 21\,cm signal with given depth, width and centre frequency across the entire sky.
Finally we convolve this map with a given antenna beam to simulate the expected power the antenna would receive per unit frequency

\subsection{Modelling The Moon}
\label{sec:moonmodel}

Though often appearing larger \citep{Hershenson1989TheIllusion}, the moon only takes up half a minute of arc across the sky \citepalias{ptolemy_almagest}\footnote{While this text was originally written circa 150, we reference the translation by G.J. Toomer in 1998}.
The GSM has 3,145,728 equal area pixels across the sky \citep{deOliveira-Costa2008AGHz} - meaning our moon only takes up \(\sim15\) pixels depending on its declination in the sky.
We therefore do not aim to model the physical variation of the temperature of the moon between pole and equator.
We also do not attempt to describe how the temperature of the lunar surface during the day varies from dawn to noon, nor at night from dusk to midnight.
Instead we take the average lunar daytime temperature to be 280\,K and the average nighttime temperature to be 50\,K \citep{Williams2017TheExperiment}.
We take a weighted average of these values based on the phase of the moon for a given time and date to give an estimate of the thermal emission over the entire surface of the moon.
We then must account for reflected solar emission.
For this we treat the sun as a blackbody of 5800\,K.
We assume an approximate lunar albedo in the Very High Frequency radio band of 0.1 \citep{Giraud1965ASurface}, multiply the solar power by this albedo and the fraction of the moon illuminated by the sun according to its phase.
Once we have appropriately scaled the reflected solar power this is added to the power arising from lunar thermal emission.

We track the position and phase of the moon using the AstroPy \citep{TheAstropyCollaboration2022ThePackage} and PyAstronomy tools \citep{Czesla2019PyA:Packages}, allowing us to appropriately scale the power output, locality and size of the moon on the sky according to time of year and antenna location.
The area of the sky occulted by the moon is masked and replaced with our approximated lunar power.

Radio frequency interference (RFI) is a large worry in 21\,cm cosmology \citep{Leeney2023BayesianMitigation, Anstey2024EnhancedReweighting}.
Earthshine, the RFI reflected from the nearside of the moon as it passes overhead, poses challenges when it comes to recovery of the redshifted 21\,cm signal \citep{Tiwari2023MeasuringInterference}.
The description of this earthshine is a complicated one that does not vary in a precisely predictable way, so can be challenging to describe computationally.

The empirical effect of earthshine has been explored by \citet{McKinley2013Low-frequencyArray}, in which the pixels passed over by the moon when at zenith see between a 50-100\,Jy increase in power across the FM (87.5-108\,MHz) band.
The extent of this injected power will vary by location - areas of the world with more broadcasting will see higher earthshine, and by day-night cycle - some broadcasters will not transmit at night, which can lower power output when the sun is down.
Thus, these specific effects are better described empirically location to location.
This paper aims to provide an overview for occultation globally, and to avoid unnecessary error we therefore do not try to generalise this RFI, instead leaving it to future experimental work.
Here we only look to deal with the moon's thermal emission, as well as the reflected solar emission to investigate the variation in foregrounds of an astronomical source due to occultation through the lunar cycle.

\subsection{Data Analysis}
\label{sec:dataanalysis}
To recover the 21\,cm signal injected into our data we use the REACH forward modelling Bayesian analysis pipeline first developed in \citet{Anstey2020AExperiments}.

This method approximates what the sky should look like and tries to fit for the parts we don't know with certainty.
It does this by dividing it into a given number of sky regions based on their spectral indices, rotates it for a given time of observation, and creates a horizon mask around the antenna for a horizon with temperature and reflection coefficient as parameters to be fit for.
It then convolves this map of the sky and horizon with a modelled beam observing the sky.
We then fit for the spectral indices, horizon parameters,  and the parameters of the added signal to give the most likely values for these parameters when compared to the antenna temperature we acquired from our simulated observation.
We do this using the the \textsc{polychord} nested sampling algorithm \citep{Handley2015PolyChord:Sampling, Handley2015Polychord:Cosmology.} utilising the likelihood described in \citet{Pattison2025GlobalConditions}.
A more complete description of this pipeline can be found in \citet{Anstey2023UseModelling} and \citet{ Pattison2025GlobalConditions}.

Throughout this paper we do not account for the moon in our foreground modelling for signal recovery to gauge the impact of not accounting for it when trying to find the 21\,cm signal.
\section{Results}
\label{sec:Results}
We aim to analyse the relation between the change to the beam weighted sky maps and the resultant error in signal recovery.
We perform this analysis for the log spiral and dipole beams described in Section \ref{sec:beams} and the forward modelling technique we discuss in Section \ref{sec:modelling}.

In Section \ref{sec:num} we do this in a more controlled environment, where the only thing changing is the size and location of the occulting object for a given observation - altering the angular radius of the body across the sky, and fixing it in various locations over the period of a simulated observation.
In Section \ref{sec:real} we discuss more realistic simulations spanning across four different locations at varying times, such that the moon and galaxy are allowed to move through various regions of the antenna beam.
Section \ref{sec:sub} investigates the possible impact of occultation from bodies smaller than the moon.

We contextualise the results in this section by drawing boundaries on what we consider to be `good' and `bad' signal recoveries.
When comparing the quality of signal recovery we look at the root mean square error, RMSE, of our recovered signal with respect to our injected one.
This RMSE refers to the root mean squared error when comparing the injected mock signal to one that we generate using the posterior averages that our Gaussian model suggests, and will give an idea of how different our recovered signal looks from the one we injected, giving an idea of the veracity of our recovery.

In simulations we have run without occultation we expect the log spiral antenna to give an RMSE from the injected signal of between 2.5-6.5\,mK, and the dipole antenna to give an RMSE of 3.5-9.5\,mK.
We consider any recovery with an RMSE of less than 10\% of the injected signal depth to be the gold standard, and any RMSE above 20\% to become increasingly spurious.

\subsection{Numerical Testing}
\label{sec:num}

In our more controlled test, we alter the empyrean disturbance due to occultation by changing the size of the occulting object and fixing it in distinct points of the sky for different simulations to maximise and minimise the impact of occultation.

For this investigation we allow the apparent diameter of our occulting body to sit between 0.50 and 1.12 degrees on the sky, giving an area on the sky that ranges between the expected lunar size, and up to five times the area what we would anticipate seeing - for a body the size of the moon this would assume apogee and perigee of \(1.56\times10^5\)km and \(3.82\times10^5\)km respectively.

For this test we simulate a 3 hour observation starting at 2025-03-25 03:00 UTC at the REACH site in the Karoo.
This time is chosen to maximise the time for which the centre of the galaxy is in the centre of the beam, and therefore maximise disruption from occultation to the observed sky.
We choose to place and fix the occulting body at a single location for the duration of each simulated observation.
This assumption is unphysical, as any occulting body would move with the sky as the earth rotates, and a more realistic simulation would likely see a greater disruption to the foregrounds than we simulate in this section - but this will provide a baseline for determining the maximum impact due to location and size of occulting body.
We place the body due east of the receiver and vary its declination below zenith, such that for a given simulation the moon will be \(\pi/64, \pi/32, \pi/16, \pi/8, \text{ or } \pi/4\) radians below zenith.
For each of these declinations we simulate an observation with occulting radius equal to \(0.25\sqrt{x}\) degrees across the sky, where we let \(x\) be equal to integers 1 through 5.
This gives us 25 simulated observations with the body at a fixed height, giving a large sample set we can draw from to quantify the relationship between occultation of the sky and quality of signal recovery.
We quantify occultation of the sky using the fractional difference between the total integrated temperature of the beam weighted sky map with and without the body as well as the fractional difference between the total integrated temperature of the non-beam weighted sky maps with and without the occultation to quantify the importance of the a body being in the centre of the galaxy, versus the two being in the centre of the beam, such that:

\begin{equation}
    \epsilon_u = \frac{\int_{t_i} T_{\text{sky}_m}(\theta, \phi, \nu, t)dtd\Omega - \int_{t_i} T_{\text{sky}_b}(\theta, \phi, \nu, t)dtd\Omega}{\int_{t_i} T_{\text{sky}_b}(\theta, \phi, \nu, t)dtd\Omega},
\label{eq:u}
\end{equation}

\noindent and

\begin{equation}
\epsilon_{bw} = 
\frac{
\left(
\begin{aligned}
&\frac{1}{4\pi}\int_0^{4\pi} D(\theta,\phi,\nu) \int_{t_i} T_{\text{sky}_m}(\theta,\phi,\nu,t) \,dt\,d\Omega \\
&\quad - \frac{1}{4\pi}\int_0^{4\pi} D(\theta,\phi,\nu) \int_{t_i} T_{\text{sky}_b}(\theta,\phi,\nu,t) \,dt\,d\Omega
\end{aligned}
\right)
}{
\frac{1}{4\pi}\int_0^{4\pi} D(\theta,\phi,\nu) \int_{t_i} T_{\text{sky}_b}(\theta,\phi,\nu,t) \,dt\,d\Omega
}
\label{eq:bw}
\end{equation}

Where \(\epsilon_u\) and \(\epsilon_{bw}\) are the unweighted, and beam-weighted error to the sky maps due to occultation; \(T_{\text{sky}_b}\) and \(T_{\text{sky}_m}\) refer to the brightness temperature of the foreground base map, and the map containing the occulting body respectively; \(\theta\) and \(\phi\) describe the spacial coordinates of our foreground maps; \(\nu\) refers to the frequency of each map; \(t\) refers to each time bin; and \(D\) describes the directivity of a given antenna beam.

\begin{figure}
    \centering
    \hspace*{-0.5cm}
    \includegraphics[width=1.15\linewidth]{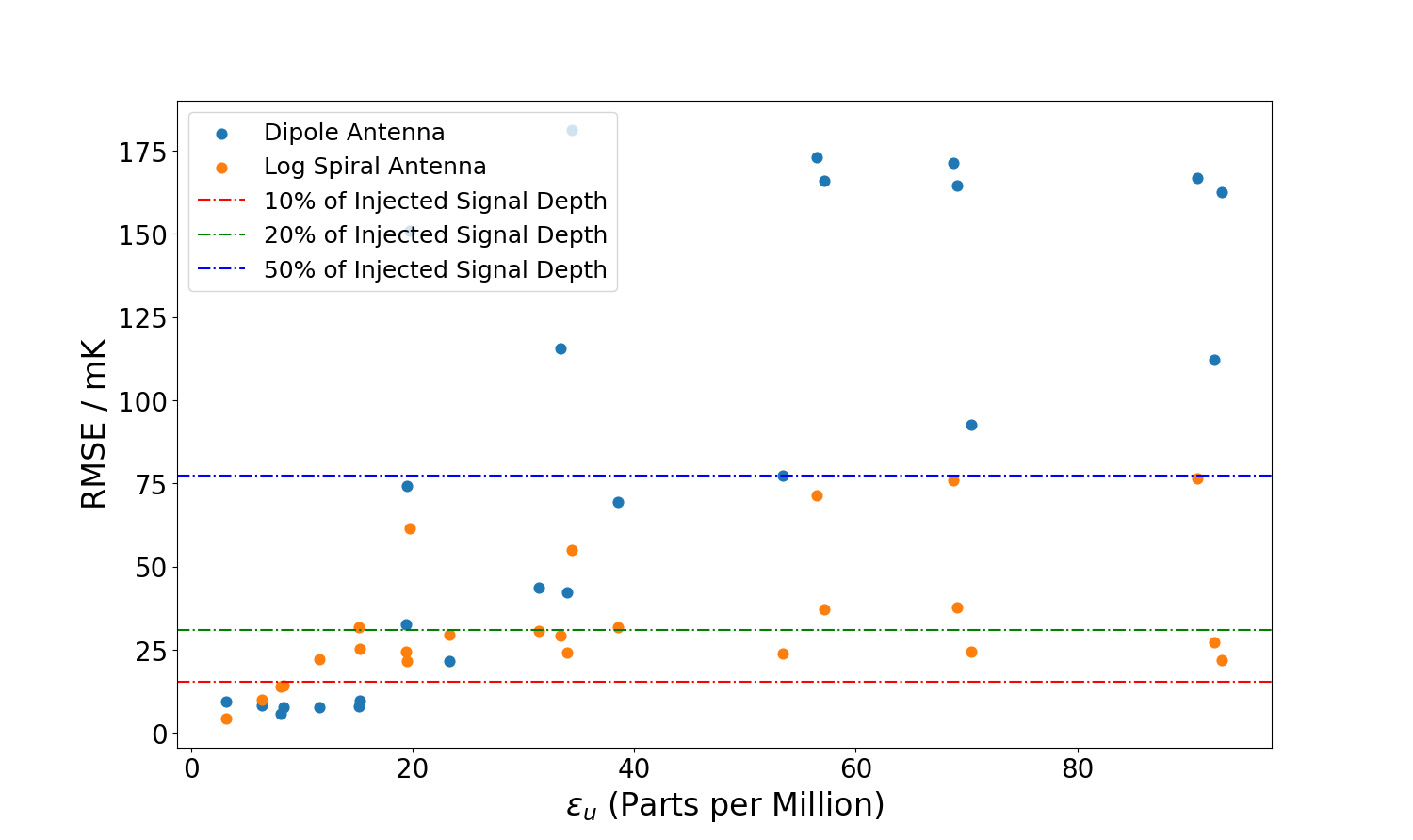}
    \caption{Comparison of  the beam-weighted error to the sky map caused by occultation and the RMSE of signal recovery for dipole and log spiral antennae.}
    \label{fig:beam}
\end{figure}

\begin{figure}
    \centering
    \hspace*{-0.5cm}
    \includegraphics[width = 1.15\linewidth]{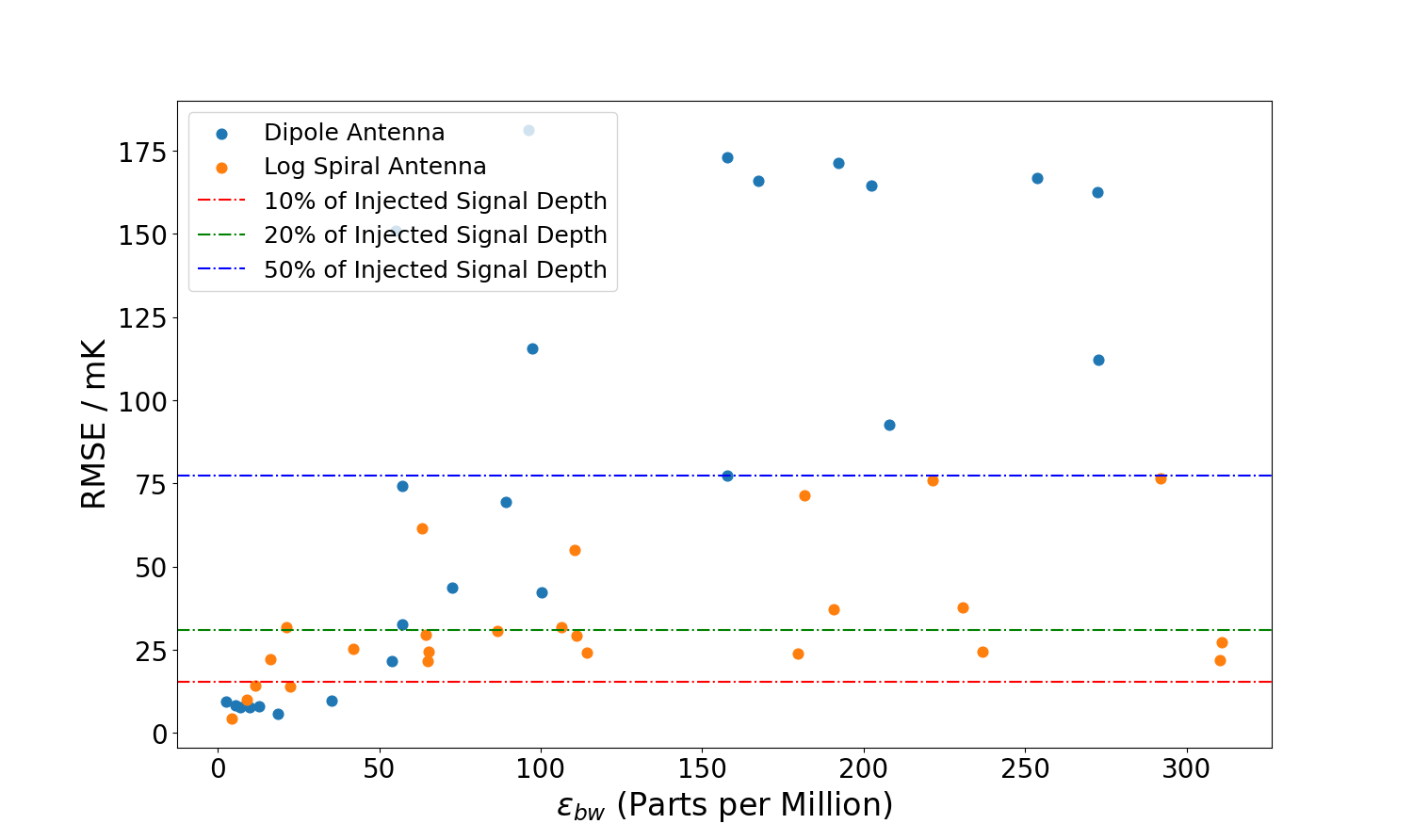}
    \caption{Comparison of the unweighted error to the sky map caused by occultation and the RMSE of signal recovery for dipole and log spiral antennae.}
    \label{fig:nobeam}
\end{figure}

We directly compare the RMSE of recovered signal with the error in the sky map due to  occultation (Figures \ref{fig:beam} and  \ref{fig:nobeam}).
We find a strong correlation between the RMSE of signal recovery and the beam weighted error in the sky map for the dipole antenna, with a Pearson correlation coefficient of \(\rho = 0.763\) - the Pearson correlation coefficient here describes the linear correlation between two data sets, it's precise definition being the ratio between the covariance of two variables over the product of their standard deviations.
The log spiral has a less direct, but still positive correlation between RMSE and \(\epsilon_{bw}\), with a \(\rho = 0.454\).
These correlations are exactly what one would anticipate from a na\"ive perspective, though the less than perfect correlation indicates more nuance than a one to one relation between the error in the beam weighted sky map and the efficacy of signal recovery.

\begin{figure}
    \centering
    \includegraphics[width=\linewidth]{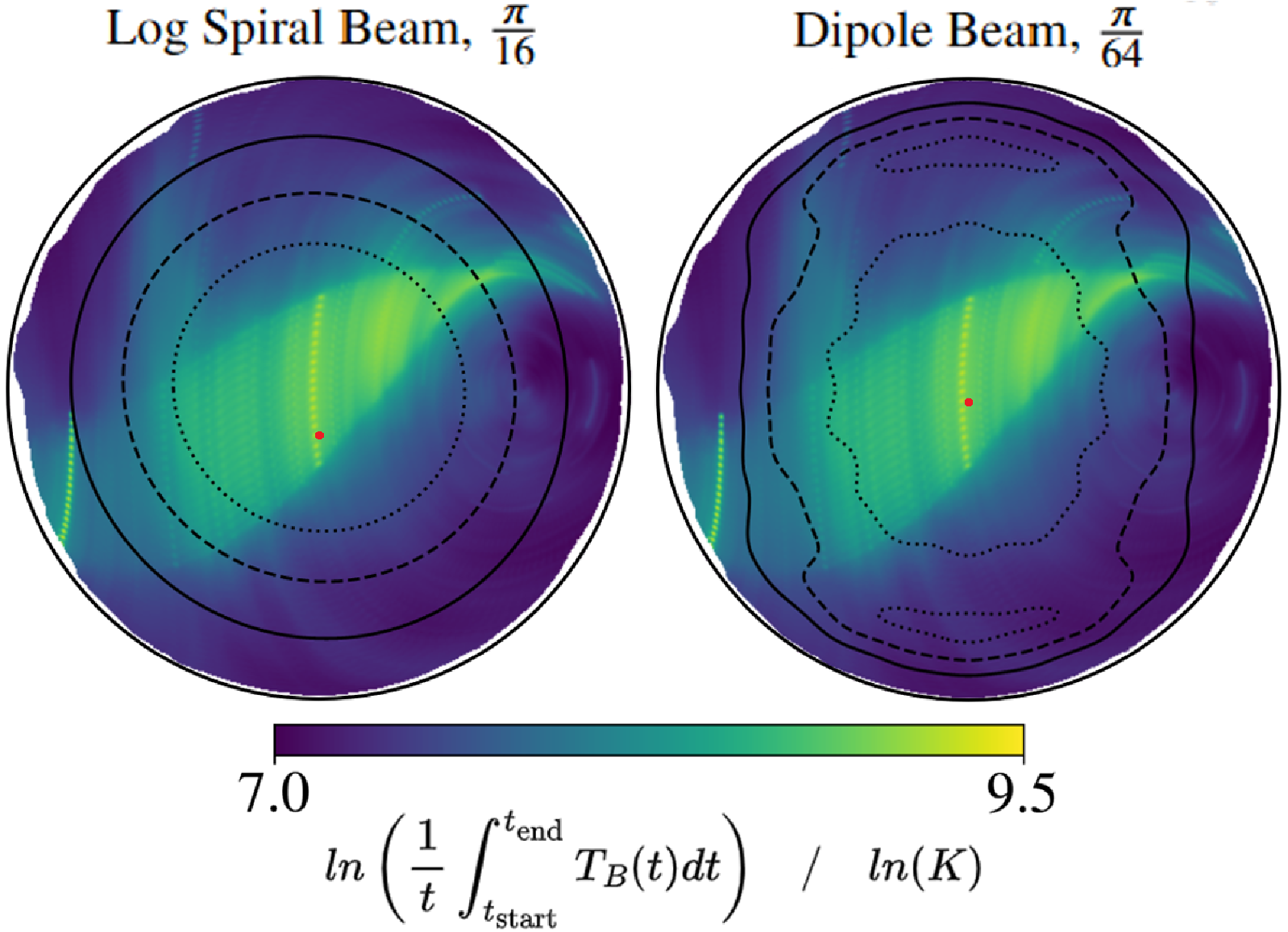}
    \caption{Plots of simulated sky maps with the moon fixed at different points in the sky during the course of the observation. This is a 3 hour observation starting at 2025-03-22 03:00:00 UTC at the REACH Karoo site. The colour map shows the natural log of the sum of the power in Kelvin at 85\,MHz of 18 ten minute observations over the course of the observation. The solid black line, the dashed black line, and the dotted black line, show the 25\%, 50\%, and 75\% power contours of the vacuum-backed log spiral (left) and dipole (right) antenna gain patterns respectively. The moon is positioned due east at \(\pi/16\) (left) and \(\pi/64\) (right) radians below zenith, and the radius of the moon is scaled up threefold and coloured red for ease of viewing. The horizon surrounding the antenna is left white, and the black circle around the horizon delineates an angular distance of 90\(^\circ\) from the zenith.}
    \label{fig:mooninsky}
\end{figure}

\begin{figure}
    \centering
    \includegraphics[width=\linewidth]{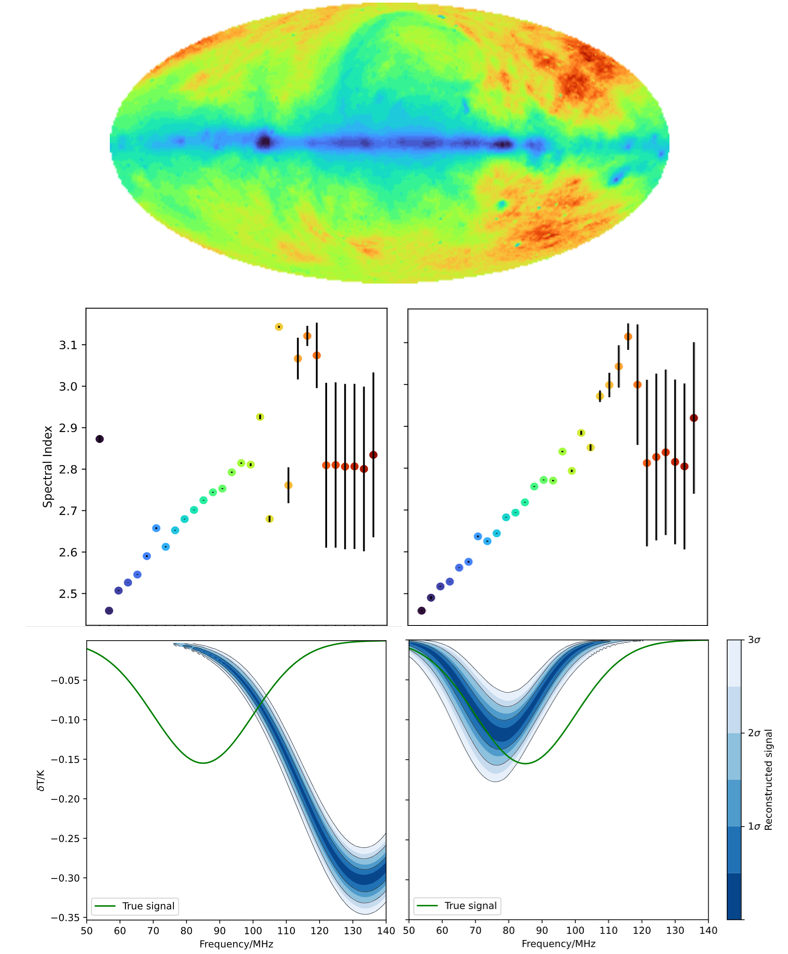}
    \caption{[Top]  Map of the entire sky in galactic coordinates. The sky is subdivided into 30 regions, where each point in a region is assumed to have the same spectral index as every other point in that region. The dark blue regions have a low spectral index, the dark red regions have a higher spectral index. [Middle] Plots of the inferred spectral indices of our 30 region model for the simulated observations where the moon is fixed at \(\pi/16\) (left) and \(\pi/64\) (right) radians below zenith. The colours of each point in the plot maps to the corresponding coloured region in the above sky map. [Bottom] Recovery of a redshifted 21\,cm signal following 3 hour mock observation using the vacuum-backed dipole starting at 2025-03-22 03:00:00 where the moon is fixed at \(\pi/16\) (left) and \(\pi/64\) (right) radians below zenith.  Injected `True’ signal shown in green, with 85 MHz  Central Frequency, 15 MHz Bandwidth, 0.155 K Depth. }
    \label{fig:specregionerrors}
\end{figure}

We can unwrap this nuance by taking two key examples, shown in Figure \ref{fig:mooninsky}, where we fix the body at \(\pi/64 \text{ and } \pi/16\) radians below zenith respectively.
The occultation being closer to the centre of the beam in the \(\pi/64\) case causes a greater overall disruption to the beam weighted sky map (57.2 parts per million) as than the \(\pi/16\) case (55.1 parts per million).
However, the \(\pi/64\) case has a better signal recovery than the \(\pi/16\) case, as we show in Figure \ref{fig:specregionerrors}.
This inconsistency between sky map error and RMSE arises due to the way the galaxy sweeps through the sky - the central galactic bulge is obscured partially in the \(\pi/16\) case, but not in the \(\pi/64\) case.
Obscuring parts of the sky with greater power will lead to a worse understanding of the most important spectral regions for our model, so despite the \(\pi/64\) case having a greater nominal impact on the beam weighted sky map, it has a smaller impact on signal recovery.
This is demonstrated in Figure \ref{fig:specregionerrors}.
These spectral regions, divided as in \citet{Anstey2020AExperiments} such that each of the N spectral regions has equal width across the total range of spectral indices, should show a monotonic increase as one moves through each region\footnote{We note that as one reaches comparatively high spectral indices the size of the regions having those spectral indices decreases and thus uncertainty on the recovered spectral index of the region increases.}.
However, we see that for the \(\pi/16\) case, the first spectral region, covering the galactic bulge, has a spectral index of almost 0.5 more than what we would anticipate. 
This leaves the pipeline unable to accurately describe region of the sky with the most power output and therefore unable to effectively create  foreground map of the sky, meaning accurate signal recovery is no longer a possibility.

This discrepancy is resolved when we directly compare the raw, unweighted, sky map error due to occultation with the RMSE of signal recovery in this case.
The \(\pi/16\) case has an unweighted sky map error from occultation of 19.7 parts per million, which sits above the \(\pi/64\) case, where the error is 19.4 parts per million.

Unfortunately, we once again cannot solely rely on the unweighted sky map error as a gauge for the impact of lunar occultation as shown in Figure \ref{fig:nobeam}.

While the dipole and log spiral antennae do have a positive correlation between the unweighted sky map error and the RMSE of signal recovery it is still not one to one.
The log spiral antenna has a \(\rho = 0.473\), and the dipole antenna gives a \(\rho = 0.750\).
This shows us that while the specific areas of the map being obscured are a large factor in the effectualness of signal recovery with our forward modelling pipeline, we cannot simply exclude the beam and its impact on the sky in quantifying the extent of the damage occultation causes to the recovery process.

We can conclude from Figures \ref{fig:beam} and \ref{fig:nobeam} that there is no exact number quantifying the disturbance to the sky map from that we can ascribe to either beam where the signal recovery process breaks down entirely.
However, we can see that at a occultation-induced beam weighted sky map error of \(\gtrsim40\), or an unweighted sky map error of \(\gtrsim15\) parts per million reliably leads to a signal recovery RMSE of 15\% or more.

From Figures \ref{fig:beam} and \ref{fig:nobeam}, we also see that the log spiral appears to be more robust to the impact of occultation than the dipole, which may be expected as the dipole is less spectrally and spatially smooth than the log spiral (Figure \ref{fig:directive}).
However, recovery using the log spiral at does reach an RMSE of 50\% of injected signal depth for varying occultations - which is a point where the signal depth priors in our Bayesian model are saturated, at which point the recovered posteriors are meaningless so we have to be careful about labelling it as robust towards occultation.
Thus, while it is true that at high levels of \(\epsilon_u\) and \(\epsilon_{bw}\) the log spiral performs better, it still reaches unacceptable levels of error.

\subsection{Realistic Simulation} 
\label{sec:real}
These numerical simulations of the sky offer some idea of how lunar disruption might impact the signal recovery process, however, the moon moves considerably over the course of the night, so more physical simulations are needed.

We select a number of dates and times across our four locations such that the moon peaks in the sky near zenith, or peaks between 40 and 70 degrees from zenith over the course of a three hour observation.
We perform these simulations for both the vacuum-backed dipole and log spiral antennae.

Across these simulations we observe a number of scenarios, some with the moon high in the sky, inside the galaxy, some with both closer to the horizon, and some with both at very different altitudes.
These simulations are all listed in Table \ref{tab:realmoon}.

\begin{table*}
\caption{Comparison of recovery of an injected redshifted 21\,cm signal with an 85\,MHz central frequency, 15\,MHz width, and 15.5\,mK signal depth in the presence of the moon at four different locations, spanning 8 observation times. \(\epsilon_u\) refers to the unweighted error to the sky map caused by lunar occultation, as described in Equation \ref{eq:u}, \(\epsilon_{bw}\) refers to the beam weighted error to the sky map caused by lunar occultation as described by Equation \ref{eq:bw}, and RMSE is the root mean square error when comparing the injected mock signal to one that we generate using the posterior averages that our Gaussian model suggests.}
\label{tab:realmoon}
\centering
\resizebox{\textwidth}{!}
{
\begin{tabular}{lccccccc}
\hline
&  Observation Start & Peak Lunar& \(\epsilon_u\) & Dipole & Dipole & Log Spiral& Log Spiral\\
&  Time & Angle &  & \(\epsilon_{bw}\)& RMSE&\(\epsilon_{bw}\) & RMSE \\
&  (UTC)&  (Degrees from Zentith)&  (Parts per Million)& (Parts per Million)& (mK)& (Parts per Million)& (mK) \\

\hline

Karoo Reserve   & 2025-03-22 03:00:00  & 4\(^\circ\) &78.2&214.0&\(50.3\pm10.2\)&250.0&\(42.5\pm4.0\)\\ 
(-30.8388N, 21.3749E)  &  2025-05-03 14:30:00 & 67\(^\circ\) & 3.9 & 7.1 &\(7.1\pm3.5\) & 4.5 & \(3.0\pm2.0\)\\
\hline
Dandiganahalli Lake & 2025-09-19 02:30:00&2\(^\circ\)& 5.2& 20.5&\(14.7\pm4.6\)&24.5&\(5.87\pm2.53\)\\
(13.9922N, 74.8760E) &2025-06-11 19:00:00  &42\(^\circ\)& 178.8&378.3&\(131.8\pm4.8\)&264.1&\(32.2\pm2.7\)\\
\hline
Inyarrimanha Ilgari Bundara & 2025-08-05 11:30:00  & \(4^\circ\) & 130.0& 357.4&  \(58.1\pm10.2\)& 416.8&\(59.2\pm2.2\)\\
(-26.7148N, 116.6044E) & 2025-11-10 19:00:00 & 56\(^\circ\)&3.4 &7.9&\(6.9\pm3.2\)&5.1&\(2.9\pm1.8\)\\
\hline
Deep Springs Valley & 2025-07-22 16:00:00 &9\(^\circ\)&8.7&31.2&\(8.6\pm2.4\)&3.6&\(5.9\pm2.6\)\\
(37.3458N, -118.0255E) & 2025-02-22 16:00:00 &67\(^\circ\)&153.0&49.0 &\(29.9\pm4.5\) &40.3&\(15.3\pm2.4\)\\
\hline

\end{tabular}
}

\end{table*}

\begin{figure}
    \centering
    \includegraphics[width=\linewidth]{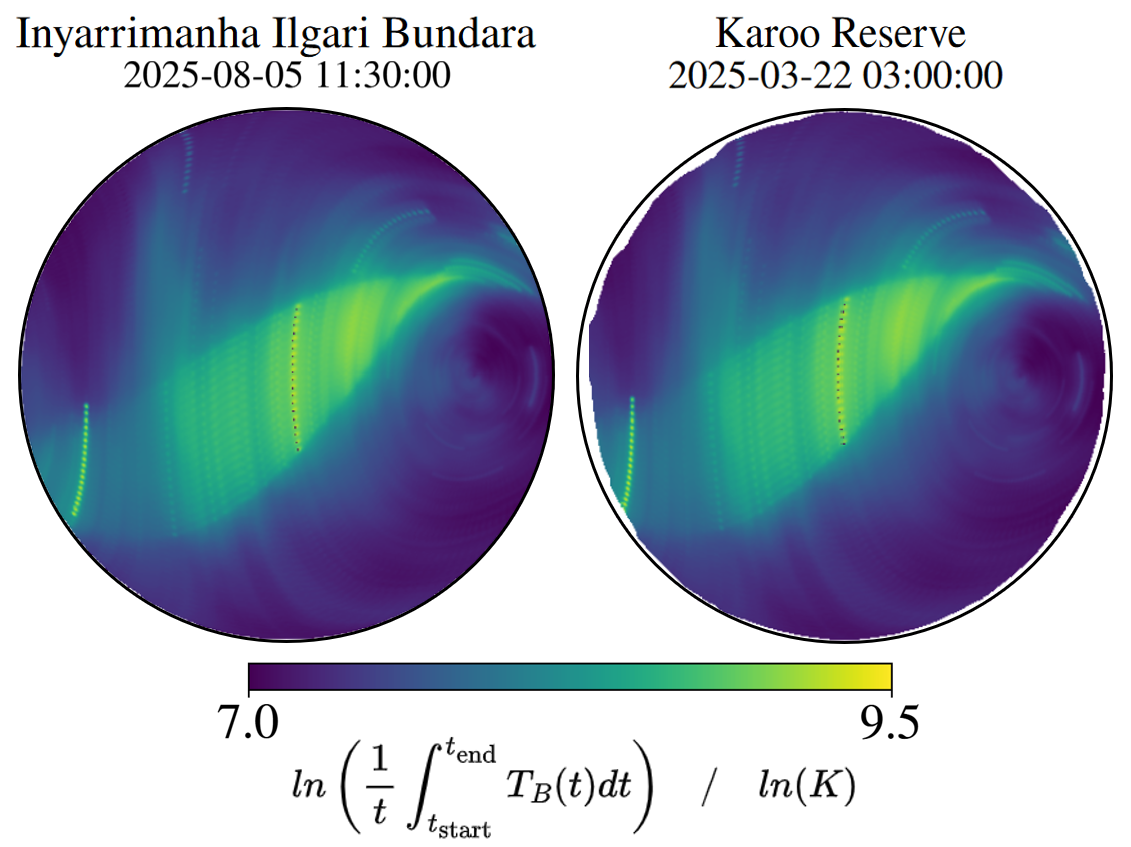}
    \caption{Simulated 3 hour observations of the sky at Inyarrimanha Ilgari Bundara starting 2025-08-05 11:30:00 UTC [left] and the REACH Karoo Site starting 2025-03-22 03:00:00 UTC [right]. The colour scale gives the log of the mean brightness temperature across the observation in the 85\,Mhz band. The horizon surrounding the antenna is left white, and the black circle around the horizon delineates an angular distance of 90\(^\circ\) from the zenith.}
    \label{fig:hghm}
\end{figure}

First, we look to the cases with the moon and galaxy both high in the sky.
This is seen for our simulated observations at the Karoo Reserve and Inyarrimanha Ilgari Bundara on the 2025-03-22 and 2025-08-05 - shown in Figure \ref{fig:hghm}.

Both of these observations had a lunar disruption to the base sky map of \(\sim\)78-130 parts per million, which crosses into the zone we established in Section \ref{sec:num} where we need to be more wary of signal recovery.
The beam-weighted sky map error caused by lunar disruption also sits between 214-357 and 250- 417 parts per million for the dipole and log spiral, which falls firmly into the region where we expect major impact to signal recovery.
In these cases RMSE in signal recovery sits between 30-38\% of the injected signal, which is an unacceptable error margin.
The values of \(\epsilon_u\) we see here are towards the upper end of what we expected to see in \ref{sec:num} - this is a result of the moon moving with the galaxy to an extent - meaning that any occultation is prolonged, unlike in the numerical case.
We can see that any observing window that would have the moon sit in the centre of the galaxy, as the galaxy reaches beam zenith must be avoided - the occultation increases the RMSE of signal recovery to an unacceptable level.

\begin{figure}
    \centering
    \includegraphics[width=\linewidth]{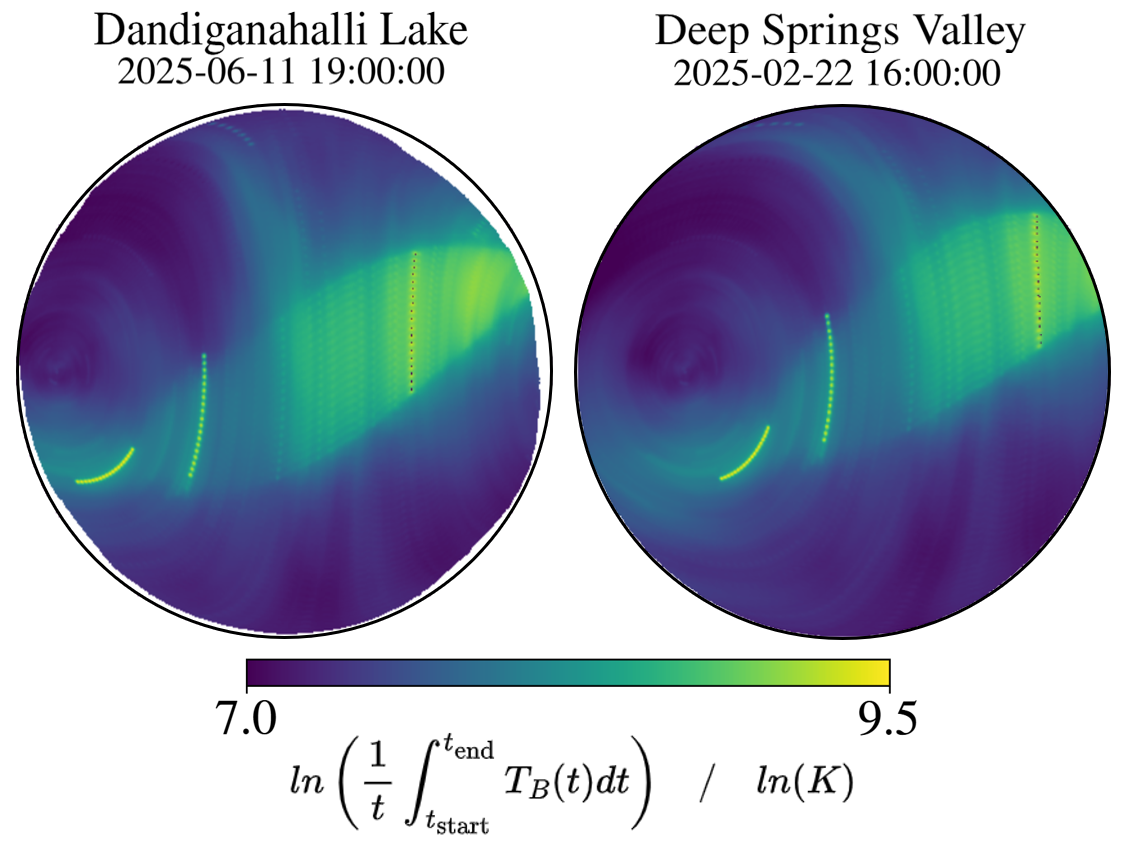}
    \caption{Simulated 3 hour observations of the sky at Dandiganahalli Lake starting 2025-06-11 19:00:00 UTC [left] and Deep Springs Valley starting 2025-02-22 16:00:00 UTC [right]. The colour scale gives the log of the mean brightness temperature across the observation in the 85\,Mhz band. The horizon surrounding the antenna is left white, and the black circle around the horizon delineates an angular distance of 90\(^\circ\) from the zenith.}
    \label{fig:sarasingal}
\end{figure}

We contrast these simulations with ones where the moon sits very close to the centre of the galaxy, but sits outside the high directivity region of the antenna beam, which focuses at zenith. 
We see this in the observations over Dandiganahalli Lake starting 2025-06-11 19:00:00 UTC and Deep Springs Valley starting 2025-02-22 16:00:00, as shown in Figure \ref{fig:sarasingal}.

These dates see the moon produce a value of \(\epsilon_u\) of 178.8 and 153.0 parts per million respectively, which are both deeply within the region for which we should get concerned with how accurately our signal is being recovered.

\begin{figure}
    \centering
    \includegraphics[width=\linewidth]{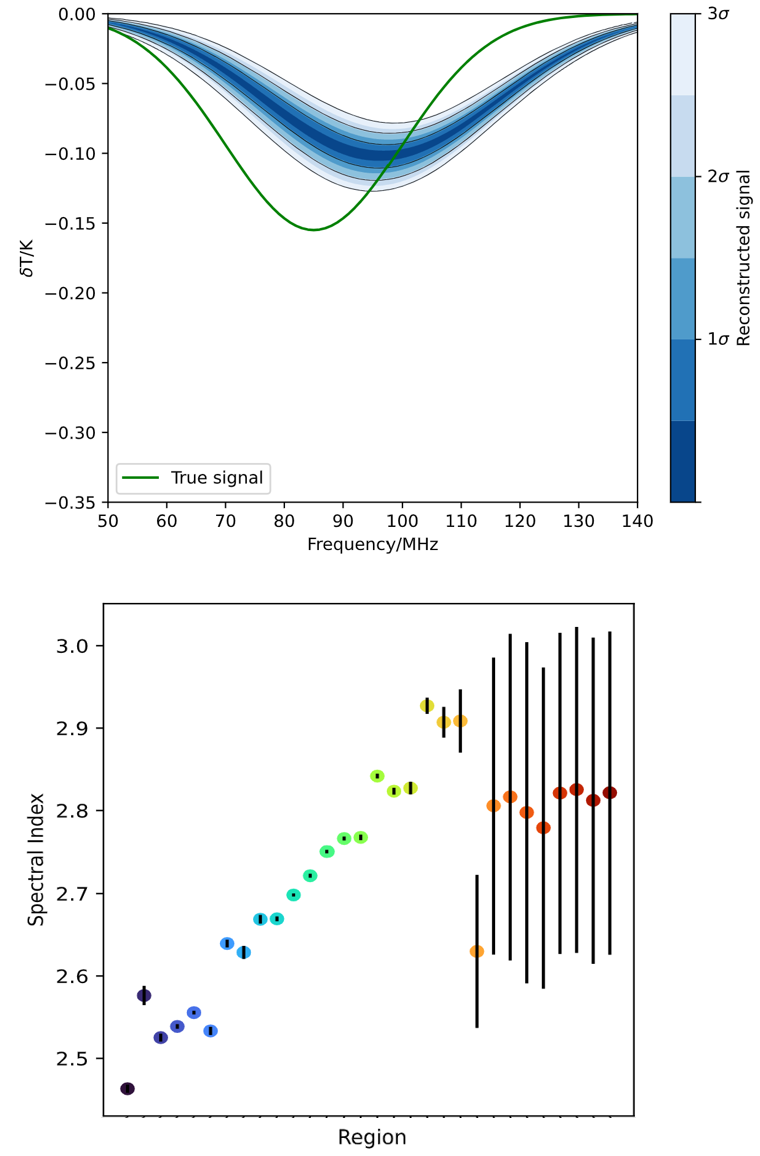}
    \caption{[Top] Recovery of a redshifted 21 cm signal following a 3 hour mock observation using the vacuum-backed log spiral antenna starting 2025-06-11 19:00:00 UTC. Injected ‘True’ signal shown in green, with 85 MHz Central Frequency, 15 MHz Bandwidth, 0.155 K Depth. [Bottom] Plot of the inferred spectral indices of a 30 region model used to describe the sky in the signal recovery. The colours of each region refer to those used for the sky map in Figure \ref{fig:specregionerrors}}
    \label{fig:sargalerrors}
\end{figure}

In the case of the Dandiganahalli Lake simulation, we observe that neither the log spiral or the dipole able to recover the signal to a level at which we are comfortable, with an RMSE of 20 and 90\% of the injected signal depth respectively.
We investigate this further in Figure
\ref{fig:sargalerrors}.
We show that the recovery using the log spiral undershoots the signal depth by about 30\%, with a slight bias towards a higher centre frequency than the injected signal.
From our plot showing the recovered spectral indices we find why we were unable to recover our signal correctly.
As in Section \ref{sec:num} we expect our plot of spectral indices to be monotonically increasing, but here we see the lower spectral index regions (near the centre of the galaxy) have clearly been corrupted by the occultation.
The date chosen sees the moon sit across a few different regions and drift slightly through the sky due to its non-sidereal motion as the observation progresses.
Because of this our pipeline cannot accurately assess the value of the spectral indices for a number of regions.
However, because we do not just see a single region corrupted, and instead see the corruption spread to a lesser degree across a number of regions, the recovery we get is slightly better than we would anticipate for such a high \(\epsilon_u\) and subsequent \(\epsilon_{bw}\).
This signal recovery does not reach our required standard for accuracy - even when the occultation happens outside of the beam zenith - but it does highlight the fact that if the occultation is spread over a number of regions, causing a slight change to the understanding of each - as opposed to entirely corrupting one - one has a better chance of decreasing error due to occultation.

We also draw attention to the simulation over Deep Springs Valley starting 2025-02-22 16:00:00 UTC.
Once again this shows a very large value of \(\epsilon_u\), at over 150 parts per million.
However, because this occultation happens so far from the peak of beam directivity, with the moon peaking at 67\(^\circ\) from zenith, the corresponding values of \(\epsilon_{bw}\) are relatively low - sitting on the cusp of an acceptable beam weighted sky map error of between 40 - 50 parts per million.
In the case of the dipole antenna here we observe an RMSE of almost 20\% of the injected signal depth, however the log spiral is able to recover the signal to an acceptable degree.
This reiterates that when trying to understand occultation we cannot only focus on where the moon is in the sky, but also where the sky is in relation to our beam.

\begin{figure}
    \centering
    \includegraphics[width=\linewidth]{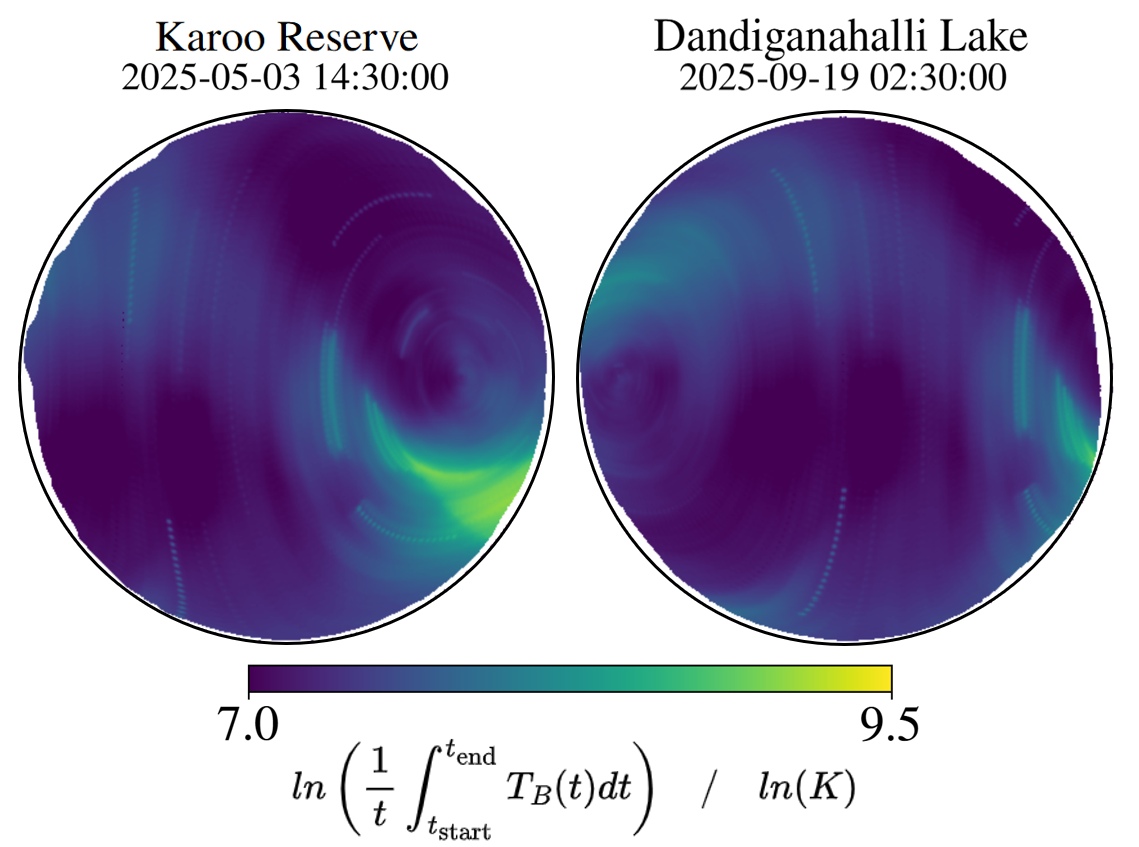}
    \caption{Simulated 3 hour observations of the sky at the REACH Karoo site starting 2025-05-03 14:30:00 UTC [left] and Dandiganahalli Lake starting 2025-09-19 02:30:00 UTC [right]. The colour scale gives the log of the mean brightness temperature across the observation in the 85\,Mhz band. The horizon surrounding the antenna is left white, and the black circle around the horizon delineates an angular distance of 90\(^\circ\) from the zenith.}
    \label{fig:hmlg}
\end{figure}

Finally we look to simulations where the moon sits near zenith, but outside of the galaxy, as in the simulated observations over the Karoo Reserve starting at 2025-05-03 14:30:00 UTC and that over Dandiganahalli Lake beginning 2025-09-10 02:30:00 UTC, shown in Figure \ref{fig:hmlg}.

We choose these dates as a representation of what happens when we have the moon occulting a relatively low power part of the sky, away from zenith, and occulting a low power part of the sky close to zenith.
For both cases we find values of \(\epsilon_u\) and \(\epsilon_{bw}\) below 6.0 and 21.0 respectively, and in both the vacuum backed log spiral and dipole antennae find the signal with an RMSE below our gold standard of 10\% injected signal depth.
This once again demonstrates the damage to signal recovery is minimal when occultation occurs outside of a high-power sky region, whether it is towards the centre of the beam or no.

For completeness we note here that the phase of the moon holds very little sway in the overall error caused by occultation.
Our largest value of \(\epsilon_u\) from occultation throughout these tests was that over Dandiganahalli lake beginning 2025-06-11 18:00:00 UTC, for which we had an \(\epsilon_u\) equal to \(\sim179\) parts per million.
This date has a lunar illumination from the sun of \(\sim99\%\).
By contrast, a new moon on this date and time would give an \(\epsilon_u\) of 180 parts per million, corresponding to a \(<1\%\) change.
This demonstrates that when discussing occultation lunar position is far more important than lunar illumination.

\subsection{Occultation from Arcminute and Sub-Arcminute Bodies}
\label{sec:sub}
In terms of apparent angular size the moon is the largest non-solar object in the sky.
However, other bodies regularly occult areas of the night sky. 
After the sun and moon the object with the largest apparent size is Venus, which has an angular diameter of approximately 1 arcminute at closest approach \citep{Galilei1613IstoriaSolari}.
We therefore investigate whether a body of this size could meaningfully impact any recovery of the 21\,cm signal.
We do note, however, that this study only looks at thermal emission and solar reflection as the only power that arises from these bodies.
Man made objects like the International Space Station (ISS), being smaller than Venus at closest approach in apparent size, is one of many satellites that communicates in the Very High Frequency (VHF) radio band that the 21-cm signal is expected to sit within, so while it, and satellites like it, will have a smaller impact on recovery due to occultation than a body like Venus, the radio frequency interference that they provide is something that must be dealt with for accurate signal recovery \citep{Leeney2023BayesianMitigation}.
Natural celestial objects, such as Jupiter, also emit non-thermally in the VHF radio band - in the case of Jovian decametric radio emission this is thought to arise from cyclotron maser instability as electrons move through the auroral regions of Jupiter's atmosphere \citep{Wu1973TheoryJupiter}.
This VHF emission may see some impact on signal recovery, but it is beyond the scope of this work.

To investigate these occultation effects we create a simulation of a worst case scenario for an object the size of Venus.
This scenario once again is a 3 hour observation  across 18 ten minute intervals starting 2025-03-25 03:00:00 UTC, where we know the galaxy passes directly over the Karoo Reserve.
We fix our Cytherian object at a Right Ascension of 17h 45m 37s and Declination of -28\(^\circ\) 56m 10s, directly in the centre of the galactic bulge \footnote{We note that this is a purely theoretical location for Venus to maximise disruption from occultation, Venus, like the moon follows closely to the ecliptic, and while it can reach declinations of \(\sim28.8 ^\circ\) \citep{Folkner2014TheDE431b} it will never perfectly occult the galactic bulge}; and we give this object the surface temperature and albedo of Venus, 736K and 0.7 \citep{1983Venus}.

Any object with the apparent size of Venus would be unresolved on a map with the resolution of the GSM.
Instead of masking the pixel containing the Cytherian object and replacing it with our new power, we instead give the occulted pixel power equal to the weighted average of the occulting body and occulted foregrounds, where the weights are determined by the fraction of the pixel taken up by the occulting body.
With an apparent size of 66 arcseconds, our occulting body takes up 7\% of a pixel, so we adjust this pixel accordingly and attempt a signal recovery.

\begin{figure}
    \centering
    \includegraphics[width=\linewidth]{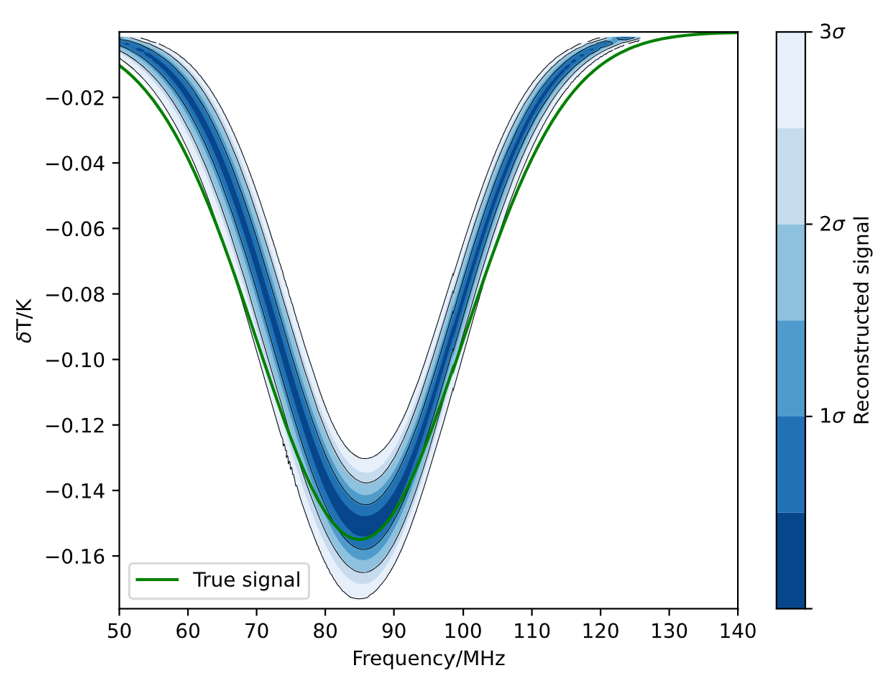}
    \caption{Recovery of a redshifted 21 cm signal following a 3 hour mock observation using the vacuum-backed dipole antenna starting 2025-03-22 03:00:00 in which a Venus-like object is placed in the centre of the galactic bulge. Injected ‘True’ signal shown in
green, with 85 MHz Central Frequency, 15 MHz Bandwidth, 0.155 K Depth.}
    \label{fig:venus}
\end{figure}

We show the recovered signal in Figure \ref{fig:venus} from the dipole antenna, where we observe that very little impact from the occulting body is observed.
The RMSE of signal recovery for the log spiral is \(13.9\pm3.9\)mK which is 9.0\% of the injected signal depth and our dipole RMSE sits at \(10.0\pm2.7\)mK - 6.5\% of injected signal.
This is anticipated, as \(\epsilon_u\) is 1.50 parts per million and \(\epsilon_{bw}\) are 4.75 and 3.58 parts per million for the log spiral and dipole antennae respectively, far below the threshold established in Section \ref{sec:num} for problematic levels of occultation.

Thus we can report that for any object that has an apparent size smaller than the moon we do not need to concern ourselves with the impact of it occulting the galactic foregrounds.

\section{Conclusions}
\label{sec:conc}

This work aims to show that occultation will be an issue for forward modelled approaches to 21\,cm cosmology.
This issue becomes increasingly prevalent as 2025 marks a major lunar standstill, in which the moon will be able to find itself in the galactic centre, causing the greatest disruption to the foregrounds we model.

We find that an error in the base sky map caused by occultation of only 15 parts per million, or a beam-weighted sky map error of between 40-50 parts per million could spike the RMSE of a signal recovery to up to 20\% of the injected signal depth, with errors beyond this in some cases leading to RMSEs of more than 50\% of injected depth - a problem that is exacerbated when the occulting object is focused into a single high-powered region of the sky.
This can be mitigated for by careful selection of observation windows - paying close attention to when the moon enters high-power regions of the sky, especially when these regions enter high directivity areas of the beam, and avoiding observing during these time periods.

We also show that bodies smaller in apparent size than the moon, the largest of which is Venus, will not impact signal recovery due to occultation, so moving observing windows to avoid planetary occultation is unnecessary.
This work does not, however, investigate the non-thermal emission from these smaller bodies in the VHF band and how these emissions could impact signal recovery - this is a question left for future research.

Throughout this work we assume a perfect understanding of the antenna beams being used for our simulation, but in reality errors in the foreground models will compound with possible error in beam modeling \citep{Cumner2024TheExperiment}, making the requirement for precision even more stringent than we detail here.
While work has been done to mitigate such errors \citep{Pagano2024AExperiments}, it is not an issue that has been solved, and will require future work to be done.

This work further exemplifies the necessity of understanding the galactic foregrounds as precisely as possible if accurate recovery of the redshifted 21\,cm signal is to be obtained.
It highlights that while lunar occultation will be a problem for 21\,cm cosmology if ignored; that with careful selection of observing windows these problems can be mitigated for and recovery of the signal, even with the moon in the sky remains an exciting possibility.

\section*{Acknowledgements}

We would like to give thanks to Will Handley for his contributions to the REACH pipeline, Oscar O'Hara for his invaluable support and Sage advice through the writing process, and to Kitty and Dylan Pattison for useful advice with respect to the tracking of spherical objects.

JHNP, DJA, and EdLA were supported by the Science and Technology Facilities Council.
We would also like to thank the Kavli Foundation for their support of REACH.

\section*{Data Availability}

The data that support the findings of this study are available from the first author upon reasonable request.



\bibliographystyle{mnras}
\bibliography{references,custom}





\bsp	
\label{lastpage}
\end{document}